\documentclass[twocolumn,aps]{revtex4}
\usepackage{graphicx}
\usepackage{amssymb}
\usepackage{amsmath}
\usepackage{bm}
\usepackage{color}

\def\e{\varepsilon}
\def\a{\alpha}
\def\b{\beta}
\def\ef{\varepsilon_{\rm F}}

\def\s{\sigma}

\def\cso{\chi^{\rm SO}}
\def\ca{\chi^{\rm intra}}
\def\ce{\chi^{\rm inter}}
\def\cs{\chi^{\rm spin}}
\def\co{\chi^{\rm orbit}}
\def\bk{{\bm k}}
\def\kd{k_{\rm D}}
\def\mb{\mu_{\rm B}}
\def\ra{\rangle}
\def\la{\langle}
\def\bo{B^{\rm orbit}}
\def\bs{B^{\rm spin}}
\def\mo{M^{\rm orbit}}
\def\ms{M^{\rm spin}}

\begin{document}

\title{Spin-orbit crossed susceptibility in topological Dirac semimetals}
\author{Yuya Ominato${}^1$, Shuta Tatsumi${}^1$, and Kentaro Nomura${}^{1,2}$}
\affiliation{${}^1$Institute for Materials Research, Tohoku University, Sendai 980-8577, Japan}
\affiliation{${}^2$Center for Spintronics Research Network, Tohoku University, Sendai 980-8577,Japan}
\date{\today}

\begin{abstract}
We theoretically study the spin-orbit crossed susceptibility of topological Dirac semimetals.
Because of strong spin-orbit coupling,
the orbital motion of electrons is modulated by Zeeman coupling, which contributes to orbital magnetization.
We find that the spin-orbit crossed susceptibility is proportional to
the separation of the Dirac points and it is highly anisotropic.
The orbital magnetization is induced only along the rotational symmetry axis.
We also study the conventional spin susceptibility.
The spin susceptibility exhibits anisotropy and the spin magnetization is induced
only along the perpendicular to the rotational symmetry axis
in contrast to the spin-orbit crossed susceptibility.
We quantitatively compare the two susceptibilities and find that they can be comparable.
\end{abstract}
\maketitle

\section{Introduction}
\label{intro}

In the presence of an external magnetic field, magnetization is induced by both the orbital motion and spin magnetic moment of electrons.
When spin-orbit coupling is negligible,
the magnetization is composed of the orbital and spin magnetization, which are induced by the minimal substitution, $\bm{p}\to\bm{p}+e\bm{A}$, and the Zeeman coupling, respectively.
Additionally, spin-orbit coupling gives rise to the spin-orbit crossed response, in which the spin magnetization is induced by the minimal substitution, and the orbital magnetization is induced by the Zeeman coupling.
In the strongly spin-orbit coupled systems, the spin-orbit crossed response can give comparable contribution to the conventional spin and orbital magnetic responses.

Spin-orbit coupling plays a key role to realize a topological phase of matter, such as topological insulators \cite{hasan2010topological} and topological semimetals \cite{armitage2018weyl}.
A natural question arising is what kind of the spin-orbit crossed response occurs in the topological materials.
Because of the topologically nontrivial electronic structure and the existence of the topological surface states, the topological materials exhibit the spin-orbit crossed response as a topological response \cite{yang2006streda,murakami2006quantum,tserkovnyak2015spin,koshino2016magnetic,nakai2016crossed}.
The spin-orbit crossed response has been investigated in several systems.
%In literary point of view, the spin-orbit crossed response in the two-dimensional (2D) quantum spin Hall insulators (QSHI) is an early example \cite{yang2006streda,murakami2006quantum}. The relation between the spin Hall conductivity and the magnetic susceptibility was discussed.
In the literature the connection between the spin-orbit crossed susceptibility and the spin Hall conductivity was pointed out \cite{yang2006streda,murakami2006quantum}.
In recent theoretical work, the spin-orbit crossed response has been investigated
%not only in the topological materials but 
also in
%strongly spin-orbit coupled systems, for example,
Rashba spin-orbit coupled systems \cite{suzuura2016theory,ando2017theory}.

The topological Dirac semimetal is one of the topological semimetals \cite{wang2012dirac,wang2013three,morimoto2014weyl,yang2014classification,pikulin2016chiral,cano2017chiral} and experimentally observed in ${\rm Na}_3{\rm Bi}$ and ${\rm Cd}_3{\rm As}_2$ \cite{liu2014discovery,neupane2014observation,borisenko2014experimental}.
The topological Dirac semimetals have an inverted band structure originating from strong spin-orbit coupling. They are characterized by a pair of Dirac points in the bulk and Fermi arcs on the surface \cite{wang2012dirac,wang2013three}. The Dirac points are protected by rotational symmetry along the axis perpendicular to the $(001)$ surface in the case of ${\rm Na}_3{\rm Bi}$ and ${\rm Cd}_3{\rm As}_2$ \cite{wang2012dirac,wang2013three}.
This is an important difference from the Dirac semimetals appearing at the phase boundary of topological insulators and ordinary insulators \cite{murakami2007phase,guo2011evolution,xu2011topological,sato2011unexpected}, in which there is no Fermi arc.
A remarkable feature of the topological Dirac semimetals is the conservation of the spin angular momentum along the rotation axis within a low energy approximation \cite{burkov2016z2}.
The topological Dirac semimetals are regarded as layers of two-dimensional (2D) quantum spin Hall insulators (QSHI) stacked in momentum space and exhibit the intrinsic semi-quantized spin Hall effect.

The magnetic responses of the generic Dirac electrons have been investigated in several theoretical papers. The orbital susceptibility logarithmically diverges and exhibits strong diamagnetism at the Dirac point \cite{fukuyama1970interband,koshino2010anomalous,koshino2016magnetic,mikitik2016magnetic}. When spin-orbit coupling is not negligible, the spin susceptibility becomes finite even at the Dirac point where the density of states vanishes \cite{koshino2016magnetic,thakur2018dynamic,zhou2018dynamical,ominato2018spin}.
This is contrast to the conventional Pauli paramagnetism and known as the Van Vleck paramagnetism
\cite{van1932theory,yu2010quantized,zhang2013topology,ominato2018spin}.

In this paper, we study the spin-orbit crossed susceptibility of the topological Dirac semimetals.
%\tcr{Recently, we published another paper where we studied the spin susceptibility of Dirac-Weyl semimetals \cite{ominato2018spin}. The current work focuses on the spin-orbit crossed susceptibility and our previous work focuses on the spin susceptibility. They are related to each other, but main interest is different.}
We find that the spin-orbit crossed susceptibility is proportional to the separation of the Dirac points and independent of the other microscopic parameters of the materials. We also include the spin conservation breaking term  which mixes up and down spins \cite{wang2012dirac,wang2013three}. We confirm that the spin-orbit crossed susceptibility is approximately proportional to the separation of the Dirac points even in the absence of the spin conservation as long as the separation is sufficiently small. We also calculate the spin susceptibility and quantitatively compare the two susceptibilities. Using the material parameters for ${\rm Na}_3{\rm Bi}$ and ${\rm Cd}_3{\rm As}_2$, we show that the contribution of the spin-orbit crossed susceptibility is important in order to appropriately estimate the total susceptibility.

The paper is organized as follows. In Sec.\ \ref{sec_model}, we introduce a model Hamiltonian and define the spin-orbit crossed susceptibility. In Secs.\ \ref{sec_so} and \ref{sec_spin}, we calculate the spin-orbit crossed susceptibility and the spin susceptibility. In Secs.\ \ref{sec_discussion} and \ref{sec_conclusion}, the discussion and conclusion are given.

\section{Model Hamiltonian}
\label{sec_model}

We consider a model Hamiltonian on the cubic lattice
\begin{align}
H_{\bk}=H_{\rm TDS}+H_{\rm xy}+H_{\rm Zeeman},
\end{align}
which is composed of three terms. The first and second terms describe the electronic states in the topological Dirac semimetals, which reduces to the low energy effective Hamiltonian around the $\Gamma$ point \cite{wang2012dirac,wang2013three,morimoto2014weyl,pikulin2016chiral,cano2017chiral}.
The first term is given by
\begin{align}
H_{\rm TDS}=\e_{\bk}+\tau_x\s_zt\sin(k_xa)-\tau_yt\sin(k_ya)+\tau_zm_\bk,
\label{eq_htds}
\end{align}
where
\begin{align}
&\e_{\bk}=C_0-C_1\cos(k_zc)-C_2\left[\cos(k_xa)+\cos(k_ya)\right], \notag \\
&m_\bk=m_0+m_1\cos(k_zc)+m_2\left[\cos(k_xa)+\cos(k_ya)\right].
\end{align}
Pauli matrices $\bm{\s}$ and $\bm{\tau}$ act on real and pseudo spin (orbital) degrees of freedom. $a$ and $c$ are the lattice constants.
$t$, $C_1$, and $C_2$ are hopping parameters. $C_0$ gives constant energy shift. $m_0$, $m_1$, and $m_2$ are related to strength of spin-orbit coupling and lead band inversion.
There are Dirac points at $(0,0,\pm k_{\rm D})$,
\begin{align}
\kd=\frac{1}{c}\arccos\left(-\frac{m_0+2m_2}{m_1}\right).
\end{align}
The separation of the Dirac points is tuned by changing the parameters, $m_0$, $m_1$, and $m_2$. The first term, $H_{\rm TDS}$, commutes with the spin operator $\s_z$, and
$H_{\rm TDS}$ is regarded as the Bernevig-Hughes-Zhang model \cite{bernevig2006quantum,morimoto2014weyl} extended to three-dimension.
The second term is given by
\begin{align}
H_{\rm xy}=&\tau_x\s_x\gamma\left[\cos(k_ya)-\cos(k_xa)\right]\sin(k_zc) \notag \\
&\hspace{0.5cm}+\tau_x\s_y\gamma\sin(k_xa)\sin(k_ya)\sin(k_zc),
\label{eq_hxy}
\end{align}
which mixes up and down spins. When $H_{\rm xy}$ is expanded around the $\Gamma$ point, leading order terms are third order terms, which are related to the rotational symmetry along the axis perpendicular to the $(001)$ surface in ${\rm Na}_3{\rm Bi}$ and ${\rm Cd}_3{\rm As}_2$.
In the current system, this axis corresponds to the $z$-axis and we call it the rotational symmetry axis in the following.
$\gamma$ corresponds to the coefficient of the third order terms in the effective model \cite{wang2012dirac,wang2013three}.
When $\gamma$ is zero, the $z$-component of spin conserves. At finite $\gamma$, on the other hand, the $z$-component of spin is not conserved.

As we mentioned in the introduction, the external magnetic field enters the Hamiltonian via the minimal substitution, $\bm{p}\to\bm{p}+e\bm{A}$, and the Zeeman coupling. We formally distinguish the magnetic field by the way it enters the Hamiltonian in order to extract the spin-orbit crossed response. $\bm{B}^{\rm orbit}$ and $\bm{B}^{\rm spin}$ represent the magnetic field in the minimal substitution and in the Zeeman coupling respectively. They are the same quantities so that we have to set $\bm{B}^{\rm orbit}=\bm{B}^{\rm spin}$ at the end of the calculation.
In the following, the subscripts $\alpha,\beta,\gamma,\delta$ refer to $x,y,z$.
We define the orbital magnetization $\mo_\a$ and the spin magnetization $\ms_\a$ as follows
\begin{align}
&\mo_\a = -\frac{1}{V}\frac{\partial \Omega}{\partial \bo_\a}, \\
&\ms_\a = -\frac{1}{V}\frac{\partial \Omega}{\partial \bs_\a},
\end{align}
where $\Omega$ is the thermodynamic potential and $V$ is the system volume.
These quantities are written, up to linear order in $\bm{B}^{\rm orbit}$ and $\bm{B}^{\rm spin}$, as
\begin{align}
&\mo_\a = \co_{\a\b}\bo_\b+\cso_{\a\b}\bs_\b, \label{eq_mo_linear} \\
&\ms_\a = \cs_{\a\b}\bs_\b+\cso_{\a\b}\bo_\b, \label{eq_ms_linear}
\end{align}
where
\begin{align}
  &\co_{\a\b}=\frac{\partial \mo_\a}{\partial \bo_\b}, \\
  &\cs_{\a\b}=\frac{\partial \ms_\a}{\partial \bs_\b}, \\
  &\cso_{\a\b}=\frac{\partial \mo_\a}{\partial \bs_\b}=\frac{\partial \ms_\a}{\partial \bo_\b}.
  \label{eq_cso_def}
\end{align}
Spin-orbit coupling can give the spin-orbit crossed susceptibility $\cso_{\a\b}$,
in addition to the conventional spin and orbital susceptibilities, $\cs_{\a\b}$ and $\co_{\a\b}$ \cite{koshino2016magnetic,nakai2016crossed}.

In the rest of the paper, we focus on the Zeeman coupling, which can induce both of the orbital and spin magnetization as we see in Eqs.\ (\ref{eq_mo_linear}) and (\ref{eq_ms_linear}).
The Zeeman coupling is given by
\begin{align}
H_{\rm Zeeman}&=-\frac{\mb}{2}
                \begin{pmatrix}
                g_s\bm{\s} & 0 \\
                0 & g_p\bm{\s}
                \end{pmatrix}
                \cdot\bm{B}^{\rm spin}, \notag \\
              &=-g_+\mb\tau_0\bm{\s}\cdot\bm{B}^{\rm spin}-g_-\mb\tau_z\bm{\s}\cdot\bm{B}^{\rm spin},
%H_{\rm Zeeman}=-\bm{M}^{\rm spin}\cdot\bm{B}^{\rm spin},
\end{align}
where $\mb$ is the Bohr magneton and $g_s,g_p$ correspond to the $g$-factors of electrons in $s$ and $p$ orbitals, respectively. We define $g_+=(g_s+g_p)/4$ and $g_-=(g_s-g_p)/4$, so that the Zeeman coupling contains two terms, the symmetric term $\tau_0\bm{\s}$ and the antisymmetric term $\tau_z\bm{\s}$ \cite{liu2010model,wakatsuki2015domain,nakai2016crossed}.
%where $\bm{M}^{\rm spin}$ is a spin magnetization operator and given by
%\begin{align}
%\bm{M}^{\rm spin}&=\frac{\mb}{2}
%                    \begin{pmatrix}
%                    g_s\bm{\s} & 0 \\
%                    0 & g_p\bm{\s}
%                    \end{pmatrix} \notag \\
%&=g_+\mb\tau_0\bm{\s}+g_-\mb\tau_z\bm{\s}.
%\end{align}

\section{Spin-orbit crossed susceptibility}
\label{sec_so}
\subsection{Formulation}
\label{sec_formula_so}
The orbital magnetization is calculated by the formula \cite{sundarm1999wave,xiao2005berry,thonhauser2005orbital,ceresoli2006orbital,shi2007quantum},
\begin{align}
M_\a^{\rm orbit}=&\frac{e}{2\hbar}\sum_n\int_{\rm BZ}\frac{d^3k}{(2\pi)^3}
                       f_{n\bk}\epsilon_{\a\beta\gamma} \notag \\
              &\times{\rm Im}\la\partial_\beta n,\bk|\left(\e_{n\bk}+H_\bk-2\mu\right)|\partial_\gamma n,\bk\ra,
\label{eq_morb}
\end{align}
where $f_{n\bk}=\left[1+e^{(\e_{n\bk}-\mu)/k_{\rm B}T}\right]^{-1}$ is the Fermi distribution function, $\partial_\a=\frac{\partial}{\partial k_\a}$, and $|n,\bk\ra$ is a eigenstate of $H_{\bk}$ and its eigenenergy is $\e_{n\bk}$.
The derivative of the eigenstates $|\partial_\a n,\bk\ra$ is expanded as \cite{ceresoli2006orbital}
\begin{align}
|\partial_\a n,\bk\ra=c_n|n,\bk\ra+\sum_{m\neq n}\frac{\la m,\bk|\hbar v_\a|n,\bk\ra}{\e_{m\bk}-\e_{n\bk}}|m,\bk\ra,
\label{eq_velocity}
\end{align}
where the velocity operator $v_\a$ is given by $v_\a=\partial_\a H_\bk/\hbar$
%\begin{align}
%v_\a=\frac{1}{\hbar}\partial_\a H_\bk,
%\end{align}
and $c_n$ is a pure imaginary number. Using Eq.\ (\ref{eq_velocity}), the formula, Eq.\ (\ref{eq_morb}), is written as
\begin{align}
\mo_\a=&\frac{e}{2\hbar}\sum_n\int_{\rm BZ}\frac{d^3k}{(2\pi)^3}f_{n\bk}\epsilon_{\a\beta\gamma} \notag \\
                     &\hspace{-1.5cm}\times{\rm Im}\sum_{m\neq n}\frac{\la n,\bk|\hbar v_\beta|m,\bk\ra\la m,\bk|\hbar v_\gamma|n,\bk\ra}{(\e_{m\bk}-\e_{n\bk})^2}(\e_{n\bk}+\e_{m\bk}-2\mu).
\label{eq_morb2}
\end{align}
We use the above formula in numerical calculation.
Using the 2D orbital magnetization $M_z^{\rm orbit(2D)}(k_z)$ at fixed $k_z$, $\mo_z$ is expressed as
\begin{align}
\mo_z=\int^{\pi/c}_{-\pi/c}\frac{d k_z}{2\pi}{M_z^{\rm orbit(2D)}}(k_z).
\end{align}
The above expression is useful when we discuss numerical results for $\cso_{zz}$.
We can relate $\chi^{\rm SO}_{\a\b}$ to the Kubo formula for the Hall conductivity,
\begin{align}
\s_{\a\beta}=&\frac{e^2}{\hbar}\sum_n\int_{\rm BZ}\frac{d^3k}{(2\pi)^3}f_{n\bk}\epsilon_{\a\beta\gamma} \notag \\
                           &\times{\rm Im}\sum_{m\neq n}\frac{\la n,\bk|\hbar v_\beta|m,\bk\ra\la m,\bk|\hbar v_\gamma|n,\bk\ra}{(\e_{m\bk}-\e_{n\bk})^2}.
\label{eq_Hall}
\end{align}
%\begin{align}
%\s_{\a\beta}=&\frac{e^2}{h}\epsilon_{\a\beta\gamma}\sum_n\int_{\rm BZ}\frac{d^3k}{(2\pi)^3}\Omega_{n\gamma}(\bm{k})f_{n\bk}
%\end{align}
%$\bm{\Omega}_n(\bm{k})=\nabla_{\bm{k}}\times\bm{A}_n(\bm{k})$, $\bm{A}_n(\bm{k})=-i\langle n,\bm{k}|\nabla_{\bm{k}}|n,\bm{k}\rangle$
When the density of states at the Fermi level vanishes, the intrinsic anomalous Hall conductivity is derived by the Streda formula \cite{streda1982theory,yang2006streda,murakami2006quantum},
\begin{align}
\s_{\a\beta}&=-e\epsilon_{\a\b\gamma}\frac{\partial \mo_\gamma}{\partial\mu}, \notag \\
&=-e\epsilon_{\a\b\gamma}\frac{\partial \cso_{\gamma\delta}}{\partial\mu}\bs_{\delta}.
\label{eq_streda}
\end{align}
The topological Dirac semimetals possess time reversal symmetry, so that the Hall conductivity is zero in the absence of the magnetic field. On the other hand, in the presence of the magnetic field, this formula suggests that the anomalous Hall conductivity at the Dirac point becomes finite beside the ordinary Hall conductivity, if $\cso_{\gamma\delta}$ is not symmetric as a function of the Fermi energy $\ef$.
In the following section, we only consider $\cso_{\a\a}$,
because $\cso_{\a\beta}$ $(\a\neq\beta)$ becomes zero from the view point of the crystalline symmetry in ${\rm Na}_3{\rm Bi}$ and ${\rm Cd}_3{\rm As}_2$.

\subsection{Numerical results}
\label{sec_num_so}
\begin{figure}
\centering
\includegraphics[width=0.95\hsize]{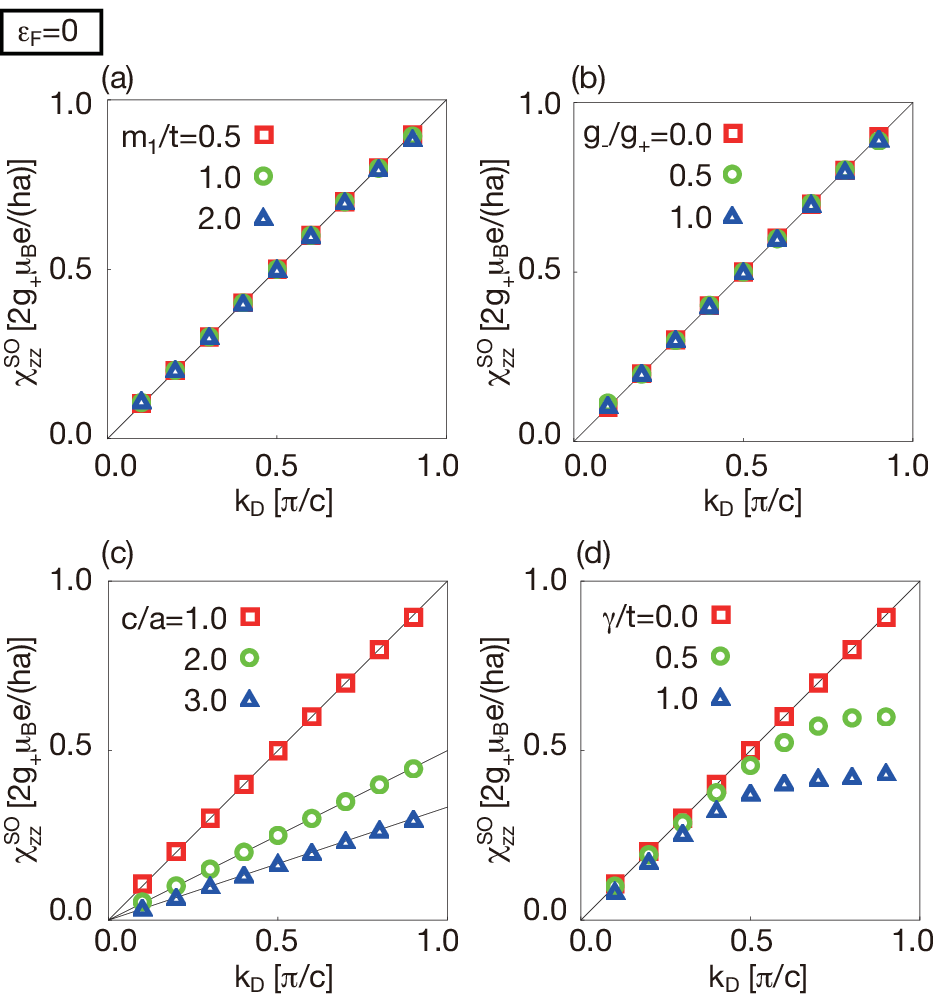}
\caption{
The spin-orbit crossed susceptibility $\cso_{zz}$ at $\ef=0$ as a function of $\kd$.
We set the parameters $m_1=m_2, m_1/t=1, g_-/g_+=1, c/a=1,$ and $\gamma=0$, if the parameters are not indicated in each figure.
The panels (a), (b), and (c) show that $\cso_{zz}$ is proportional to $\kd$, which means that $\cso_{zz}$ reflects the topological property of the electronic structure. From these numerical results, we obtain analytical expression for $\cso_{zz}$, Eq.\ (\ref{eq_cso}), which is independent of model parameters except for $\kd$ and $g_+$. The panel (d) show that $H_{\rm xy}$ reduces $\cso_{zz}$ but it is negligible for sufficiently small $\kd$.
}

\label{fig_kd_dep}
\end{figure}

Numerically differentiating Eq.\ (\ref{eq_morb2}) with respect to $\bs_\a$, we obtain $\chi^{\rm SO}_{\a\a}$.
In Sec.\ \ref{sec_so} and \ref{sec_spin}, we omit $\e_{\bk}$ in Eq.\ (\ref{eq_htds}) for simplicity.
This simplification does not change essential results in the following calculations.
In Sec. \ref{sec_discussion}, we incorporate $\e_{\bk}$ in order to compare the spin-orbit crossed susceptibility and the spin susceptibility quantitatively in ${\rm Na}_3{\rm Bi}$ and ${\rm Cd}_3{\rm As}_2$.
Figure \ref{fig_kd_dep} shows the spin-orbit crossed susceptibility $\cso_{zz}$ at $\ef=0$ as a function of the separation of the Dirac points $k_{\rm D}$. In the present model, there are several parameters, such as $t, a, m_0,$ and so on. We systematically change them and find which parameter affect the value of $\cso_{zz}$.
Figure \ref{fig_kd_dep} (a), (b), and (c) show that $\cso_{zz}$ increases linearly with $k_{\rm D}$ and satisfy following relation,
\begin{align}
	\cso_{zz}=g_+\mb\frac{2e}{h}\frac{\kd}{\pi}.
	\label{eq_cso}
\end{align}
$\cso_{zz}$ is proportional to the separation of the Dirac points $\kd$ and the coupling constant $g_+\mb$.

Eq.\ (\ref{eq_cso}) is given by numerical calculation. This result is understood as follows.
$\cso_{zz}$ is obtained as
\begin{align}
	\cso_{zz}=\int^{\pi/c}_{-\pi/c}\frac{d k_z}{2\pi}{\cso_{zz}}^{\rm (2D)}(k_z),
	\label{eq_cso2d}
\end{align}
where ${\cso_{zz}}^{\rm (2D)}(k_z)$ is the ${\rm 2D}$ spin-orbit crossed susceptibility at fixed $k_z$, which is defined in the same way as Eq.\ (\ref{eq_cso_def}).
${\cso_{zz}}^{\rm (2D)}$ is quantized as $2g_+\mb e/h$ in the 2D-QSHI and vanishes in the ordinary insulators \cite{murakami2006quantum,nakai2016crossed}.
The topological Dirac semimetal is regarded as layers of the 2D-QSHI stacked in the momentum space and the spin Chern number on the $k_x$-$k_y$ plane with fixed $k_z$ becomes finite only between the Dirac points. As a result, we obtain Eq.\ (\ref{eq_cso}).
The sign of $\cso_{zz}$ depends on the spin Chern number on the $k_x$-$k_y$ plane with fixed $k_z$ between the Dirac points.
This is analogous to the anomalous Hall conductivity in the Weyl semimetals \cite{burkov2011weyl,armitage2018weyl,burkov2016z2}. In Fig.\ \ref{fig_kd_dep} (d), $\cso_{zz}$ increases linearly at small $\kd$ but deviates from Eq.\ (\ref{eq_cso}) for finite $\gamma$.
This is because the $z$-component of spin is not conserved in the presence of $H_{\rm xy}$, Eq.\ (\ref{eq_hxy}), and the above argument for 2D-QSHI is not applicable to the present system.
In the following calculation, we set $m_0=-2m_2$, $m_1=m_2$, $m_1/t=1$ and $c/a=1$.

\begin{figure}
\centering
\includegraphics[width=1\hsize]{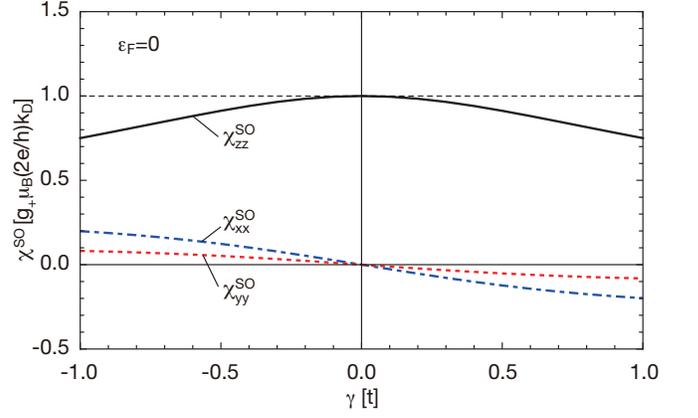}
\caption{The spin-orbit crossed susceptibility as a function of $\gamma$. The solid black curve is $\cso_{zz}$, the blue dashed curve is $\cso_{xx}$, and the red dashed curve is $\cso_{yy}$. We set the parameters $m_0=-2m_2, m_1=m_2, m_1/t=1, g_-/g_+=1$, and $c/a=1$.
Breaking the conservation of $\s_z$, i.e., with the increase of $\gamma$, $\cso_{zz}$ is reduced, while $\cso_{xx}$ and $\cso_{yy}$ become finite.
}
\label{fig_g_dep}
\end{figure}

Figure \ref{fig_g_dep} shows $\cso_{\a\a}$ at $\ef=0$ as a function of $\gamma$. At $\gamma=0$, $\cso_{zz}$ is finite as we mentioned above. On the other hand, $\cso_{xx}$ and $\cso_{yy}$ are zero. This means that the orbital magnetization is induced only along z-axis, which is the rotational symmetry axis.
As a function of $\gamma$, $\cso_{zz}$ is an even function and $\cso_{xx(yy)}$ is an odd function.

\begin{figure}
\centering
\includegraphics[width=0.95\hsize]{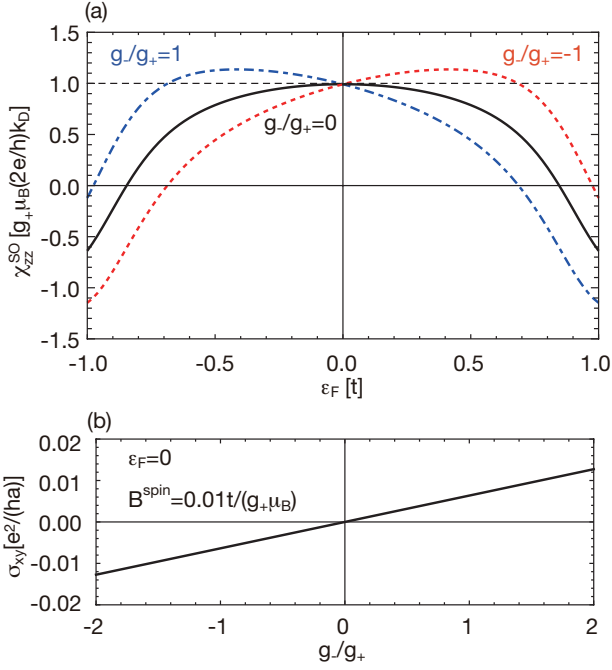}
\caption{The spin-orbit crossed susceptibility $\cso_{zz}$ as a function of $\ef$ and the Hall conductivity as a function of $g_-/g_+$. We set the parameters $m_0=-2m_2, m_1=m_2, m_1/t=1, c/a=1,$ and $\gamma=0$.
At $\ef=0$, the value of $\cso_{zz}$ is independent of $g_-$ but its $\ef$ dependence changes at finite $g_-$. Consequently, the Hall conductivity becomes finite in accordance with Eq.\ (\ref{eq_streda}).
}
\label{fig_ef_dep}
\end{figure}

Figure \ref{fig_ef_dep} (a) shows $\cso_{zz}$ around the Dirac point as a function of $\ef$. When $g_-/g_+=0$, $\cso_{zz}$ is an even function around the Dirac point. At $\ef=0$, $\cso_{zz}$ is independent of $g_-/g_+$ as we see it in Fig.\ \ref{fig_kd_dep} (b). When $g_-/g_+\neq0$, however, $\cso_{zz}$ is asymmetric and the derivative of $\cso_{zz}$ is finite.
%The sign of the derivative corresponds to the sign of $g_-/g_+$.
This suggests that the Hall conductivity is finite when $g_-/g_+\neq0$.
Calculating Eq.\ (\ref{eq_Hall}) numerically, We confirm that the Hall conductivity is finite at $\ef=0$. Figure \ref{fig_ef_dep} (b) shows $\s_{xy}$ as a function $g_-/g_+$. $\s_{xy}$ linearly increases with $g_-/g_+$. The topological Dirac semimetal is viewed as a time reversal pair of the Weyl semimetal with up and down spin.
Therefore, the Hall conductivity completely cancel with each other. Even in the presence of $g_+$ Zeeman term (the symmetric term), the cancellation is retained. In the presence of $g_-$ Zeeman term (the antisymmetric term), on the other hand, the cancellation is broken. This is because $g_-$ Zeeman term changes the separation of the Dirac points and the direction of the change is opposite for the up and down spin Weyl semimetals.
As a result, the Hall conductivity is finite in $g_-/g_+\neq0$ and given by
\begin{align}
\s_{xy}=\frac{2}{\pi}\frac{e^2}{ha}\frac{g_-\mb B^{\rm spin}}{t}.
\end{align}
This expression is quantitatively consistent with the numerical result in Fig.\ \ref{fig_ef_dep} (b).

%The Hamiltonian of topological Dirac semimetals is block diagonal and written as
%\begin{align}
%H_{\rm TDS}=\begin{pmatrix}
%                       H_{\rm WS}^+ & 0 \\
%                       0 & H_{\rm WS}^-
%                      \end{pmatrix}.
%\end{align}

%The Hall conductivity is given by
%\begin{align}
%\s_{\rm AHE}^s=s\frac{e^2}{h}\frac{\kd+s\delta k}{\pi}
%\end{align}
%where $s=\pm1$, $\delta k=g_-\mb B^{\rm spin}/(ta)$.
%Charge density and Hall conductivity is related as
%\begin{align}
%\rho^s=\s_{\rm AHE}^sB_z^{\rm orbit}.
%\end{align}
%The expectation value of spin is calculated as
%\begin{align}
%\la\s_z\ra&=\frac{1}{e}(\rho^+-\rho^-) \notag \\
%%&=\frac{1}{e}(\s_{\rm AHE}^+-\s_{\rm AHE}^-)B_z^{\rm orbit} \notag \\
%&=\frac{2e}{h}\frac{\kd}{\pi} B_z^{\rm orbit}
%\end{align}
%The spin magnetization is given by
%\begin{align}
%\la M_z^{\rm spin}\ra&=g_+\mb\la\s_z\ra \notag \\
%&=g_+\mb\frac{2e}{h}\frac{\kd}{\pi} B_z^{\rm orbit}
%\end{align}

\section{Spin susceptibility}
\label{sec_spin}

In this section, we calculate the spin susceptibility using the Kubo formula,
\begin{align}
\chi^{\rm spin}_{\a\a}(\bm{q},\ef)=&\frac{1}{V}\sum_{nm\bm{k}}\frac{-f_{n\bm{k}}+f_{m\bm{k}-\bm{q}}}{\e_{n\bm{k}}-\e_{m\bm{k}-\bm{q}}} \notag \\
&\times\mb^2\left|\la n,\bk|g_+\tau_0\s_\a+g_-\tau_z\s_\a|m,\bk-\bm{q}\ra\right|^2,
\label{eq_kubo_cs}
\end{align}
where $V$ is the system volume, $f_{n\bm{k}}$ is the Fermi distribution function,
$\e_{n\bm{k}}$ is energy of $n$-th band and $|n,\bm{k}\ra$ is a Bloch state of the unperturbed Hamiltonian.
Taking the long wavelength limit $|\bm{q}|\to0$, we obtain 
\begin{align}
\lim_{|\bm{q}|\to0}\chi^{\rm spin}_{\a\a}(\bm{q},\ef)=\ca_{\a\a}(\ef)+\ce_{\a\a}(\ef),
\end{align}
where $\ca_{\a\a}(\ef)$ is an intraband contribution,
\begin{align}
\ca_{\a\a}(\ef)=&\frac{1}{V}\sum_{n\bm{k}}\left(-\frac{\partial f_{n\bm{k}}}{\partial\e_{n\bm{k}}}\right) \notag \\
&\times\mb^2\left|\la n,\bm{k}|g_+\tau_0\s_\a+g_-\tau_z\s_\a|n,\bm{k}\ra\right|^2,
\label{eq_ca}
\end{align}
and $\ce_{\a\a}(\ef)$ is an interband contribution,
\begin{align}
\ce_{\a\a}(\ef)=&\frac{1}{V}\sum_{n\neq m,\bm{k}}\frac{-f_{n\bk}+f_{m\bk}}{\e_{n\bk}-\e_{m\bk}} \notag \\
&\times\mb^2\left|\la n,\bm{k}|g_+\tau_0\s_\a+g_-\tau_z\s_\a|m,\bm{k}\ra\right|^2.
\label{eq_ce}
\end{align}
At the zero temperature, only electronic states on the Fermi surface contribute to $\ca_{\a\a}$. On the other hand, all electronic states below the Fermi energy can contribute to $\ce_{\a\a}$ \cite{ominato2018spin}.
From the above expression, we see that $\ce_{\a\a}$ becomes finite, when the matrix elements of the spin magnetization operator between the conduction and valence bands is non-zero, i.e. the commutation relation between the Hamiltonian and the spin magnetization operator is non-zero.
If the Hamiltonian and the spin magnetization operator commute,
\begin{align}
\la n,\bm{k}|\left[H_\bk,g_+\tau_0\s_\a+g_-\tau_z\s_\a\right]|m,\bm{k}\ra=0,
\end{align}
the interband matrix element satisfies
\begin{align}
(\e_{n\bk}-\e_{m\bk})\la n,\bm{k}|g_+\tau_0\s_\a+g_-\tau_z\s_\a|m,\bm{k}\ra=0.
\end{align}
This equation means that there is no interband matrix element and $\ce_{\a\a}=0$, because $\e_{n\bk}-\e_{m\bk}\neq0$.

\begin{figure}
\begin{center}
\includegraphics[width=1\hsize]{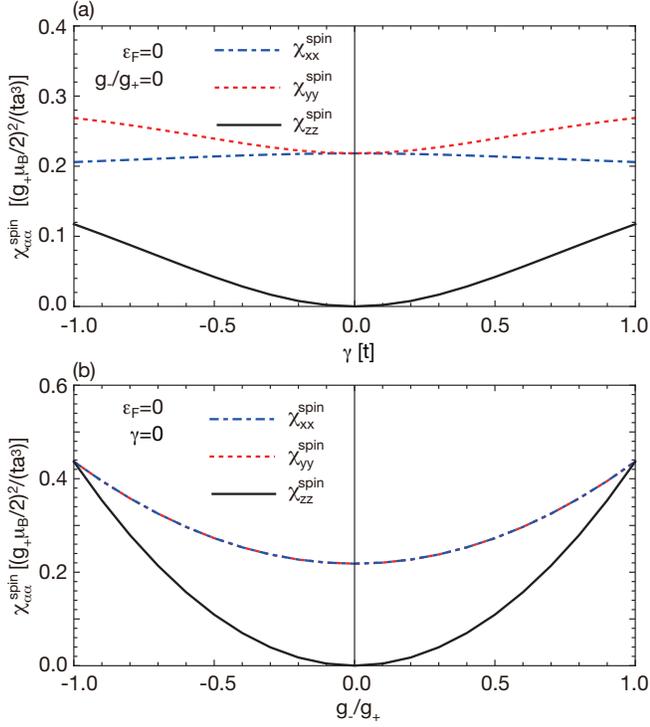}
\end{center}
\caption{The spin susceptibility $\cs_{\a\a}$ at $\ef=0$ as a function of (a) $\gamma$ and (b) $g_-/g_+$. We set $m_0=-2m_2$, $m_1=m_2$, $m_1/t=1$, and $c/a=1$.
At $\gamma=0$ and $g_-/g_+=0$, $\cs_{zz}=0$ while $\cs_{xx},\cs_{yy}>0$. These behaviors are explained by the commutation relation between the Hamiltonian and the spin magnetization operators as discussed in the main text.
}
\label{fig_spin}
\end{figure}

In the following, we set $\ef=0$, where the density of states vanishes. Therefore, there is no intraband contribution and we only consider the interband contribution. We numerically calculate Eq.\ (\ref{eq_ce}).
Figure \ref{fig_spin} shows the spin susceptibility $\cs_{\a\a}$ as a function of (a) $\gamma$ and (b) $g_-/g_+$.
%At $\ef=0$, the Density of states vanishes, so that $\ca_{\a\a}$ gives no contribution and $\cs_{\a\a}$ is solely given by $\ce_{\a\a}$.
%When the matrix elements of the spin operator between the conduction and valence bands is non-zero,
%$\ce_{\a\a}$ becomes finite.
In the following, we explain the qualitative behavior of $\cs_{\a\a}$ using the commutation relation between the Hamiltonian and the spin magnetization operator.
In Fig.\ (\ref{fig_spin}) (a), $\cs_{zz}$ vanishes at $\gamma=0$, because the Hamiltonian, $H_{\rm TDS}$, and the spin magnetization operator of $z$-component, $g_+\mb\tau_0\s_z$, commute,
%In both figures, $\cs_{xx}$ and $\cs_{yy}$ are finite, but $\cso_{zz}$ vanishes at $\gamma=0$ and $g_-=0$. These behavior are explained by the following relations,
%\tcr{
%\begin{align}
%0&=\la n,\bm{k}|[H_{\rm TDS},g_+\mb\tau_0\s_z]|m,\bm{k}\ra \notag \\
%&=(\e_{n\bk}-\e_{m\bk})\la n,\bm{k}|g_+\mb\tau_0\s_z|m,\bm{k}\ra.
%\end{align}
%}
\begin{align}
[H_{\rm TDS},g_+\mb\tau_0\s_z]=0.
\end{align}
For finite $\gamma$, on the other hand, $\cs_{zz}$ increases with $|\gamma|$. This is because the commutation relation between $H_{\rm xy}$ and $g_+\mb\tau_0\s_z$ is non-zero,
\begin{align}
[H_{\rm xy},g_+\mb\tau_0\s_z]\neq0,
\end{align}
and $\ce_{zz}$ gives finite contribution.
$\cs_{xx}$ and $\cs_{yy}$ are finite even in the absence of $H_{\rm xy}$, i.e. $\gamma=0$, because $H_{\rm TDS}$ and $g_+\mb\tau_0\s_\a {~}(\a=x,y)$ do not commute,
\begin{align}
[H_{\rm TDS},g_+\mb\tau_0\s_x]\neq0, \notag \\
[H_{\rm TDS},g_+\mb\tau_0\s_y]\neq0.
\end{align}
At $\gamma=0$, $\cs_{xx}$ is equal to $\cs_{yy}$. For finite $\gamma$, however, they deviate from each other. This is because $H_{\rm TDS}$ possesses four-fold rotational symmetry along $z$-axis but $H_{\rm xy}$ breaks the four-fold rotational symmetry.
Figure (\ref{fig_spin}) (b) shows that $\cso_{zz}$ becomes finite when $g_-/g_+\neq0$.
The antisymmetric term, $g_-\mb\tau_z\s_z$, and $H_{\rm TDS}$ do not commute,
\begin{align}
[H_{\rm TDS},g_-\mb\tau_z\s_z]\neq0.
\end{align}
Consequently, $\ce_{zz}$ gives finite contribution, though the $z$-component of spin is a good quantum number.
The antisymmetric term does not break the four-fold rotational symmetry along $z$-axis, so that $\cs_{xx}$ is equal to $\cs_{yy}$ in Fig.\ (\ref{fig_spin}) (b).

The spin susceptibility $\cs_{\a\a}$ is also anisotropic but contrasts with the spin-orbit crossed susceptivity $\cso_{\a\a}$. $\cs_{xx}$ and $\cs_{yy}$ are larger than $\cs_{zz}$, in contrast $\cso_{zz}$ is larger than $\cso_{xx}$ and $\cso_{yy}$. Therefore, the angle dependence measurement of magnetization will be useful to separate the contribution from the each susceptibility.

\section{Discussion}
\label{sec_discussion}
In this section, we quantitatively compare the spin-orbit crossed susceptibility $\cso_{zz}$ and the spin susceptibility $\cs_{zz}$ at the Dirac points as a function of $g_-/g_+$. In the following calculation, we set the parameters to reproduce the energy band structure around the $\Gamma$ point calculated by the first principle calculation for ${\rm Cd}_2{\rm As}_3$ and ${\rm Na}_3{\rm Bi}$ \cite{wang2012dirac,cano2017chiral}. The parameters are listed in the table and we omit $H_{\rm xy}$, i.e. $\gamma=0$.
%The parameters in the lattice model are set as $c_0=0.306, c_1=0.033, c_2=0.144, m_0=0.376, m_1=-0.058, m_2=-0.169, t=0.070$ and $a=12.64, c = 25.43$, which are the parameters for ${\rm Cd}_3{\rm As}_2$ \cite{liu2014discovery}. $c_0=-1.183, c_1=0.188, c_2=-0.654, m_0=1.754, m_1=-0.228, m_2=-0.806, t=0.485$ and $a=5.07, c= 9.66$.

\begin{figure}[t]
\begin{center}
\leavevmode\includegraphics[width=1\hsize]{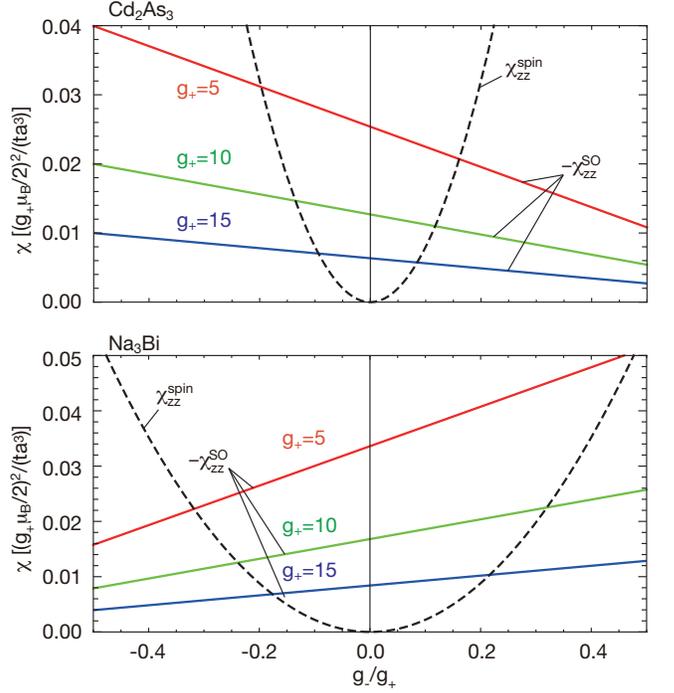}
\end{center}
\caption{The spin-orbit crossed susceptibility $\cso_{zz}$ and the spin susceptibility $\cs_{zz}$ at the Dirac points as a function of $g_-/g_+$. The dashed curve is $\cs_{zz}$ and the solid lines are $\cso_{zz}$. The upper (lower) panel shows ${\rm Cd}_2{\rm As}_3$ (${\rm Na}_3{\rm Bi}$).
When $g_-/g_+$ are sufficiently small, $\cso_{zz}$ becomes comparable to $\cs_{zz}$.
}
\label{fig_comp}
\end{figure}

Figure \ref{fig_comp} shows the two susceptibilities as a function of $g_-/g_+$. We find that the two susceptibilities are approximately written as 
\begin{align}
\chi^{\rm spin}_{zz}\sim\left(\frac{g_-}{g_+}\right)^2,
\end{align}
and
\begin{align}
\cso_{zz}\sim-\frac{1}{g_+}\left(\chi_0+\frac{g_-}{g_+}\right),
\end{align}
by numerical fitting.
In the present parameters, $\cso_{zz}$ is negative and depends on $g_-/g_+$. The dependence on $g_-/g_+$ originates from the existence of $\e_{\bk}$, which breaks the particle-hole symmetry. The $g$-factors are experimentally estimated as $g_s=18.6$ for ${\rm Cd}_2{\rm As}_3$ \cite{jeon2014landau} and $g_-=20$ for ${\rm Na}_3{\rm Bi}$ \cite{xiong2015evidence}. Unfortunately, there is no experimental data which determines both of $g_s,g_p$ or $g_+,g_-$. From Fig.\ \ref{fig_comp},
we see that $\cso_{zz}$ can dominate over $\cs_{zz}$ if $g_-/g_+\simeq0$.
As far as we know, there is no experimental observation of the magnetic susceptibility in these materials. We expect the experimental observation in near future and our estimation of $\cso_{zz}$ will be useful to appropriately analyze experimental data.

\begin{center}
  \begin{tabular}{|l|r|r|} \hline
    \multicolumn{3}{|c|}{Material parameters} \\ \hline
     & ${\rm Cd_3As_2}$ & ${\rm Na_3Bi}$ \\ \hline
     $C_0$ & 0.306[eV] & -1.183[eV] \\ \hline
     $C_1$ & 0.033[eV] & 0.188[eV] \\ \hline
     $C_2$ & 0.144[eV] & -0.654[eV] \\ \hline
     $m_0$ & 0.376[eV] & 1.754[eV] \\ \hline
     $m_1$ & -0.058[eV] & -0.228[eV] \\ \hline
     $m_2$ & -0.169[eV] & -0.806[eV] \\ \hline
     $t$ & 0.070[eV] & 0.485[eV] \\ \hline
     $a$ & 12.64[$\AA$] & 5.07[$\AA$] \\ \hline
     $c$ & 25.43[$\AA$] & 9.66[$\AA$] \\ \hline
  \end{tabular}
\end{center}

%\begin{figure}
%\begin{center}
%\leavevmode\includegraphics[width=0.9\hsize]{fig_compNaBi.eps}
%\end{center}
%\caption{The spin-orbital susceptibility and the spin susceptibility as a function of $g_-/g_+$.
%}
%\label{fig_comp}
%\end{figure}

\section{Conclusion}
\label{sec_conclusion}
We theoretically study the spin-orbit crossed susceptibility of topological Dirac semimetals. We find that the spin-orbit crossed susceptibility along rotational symmetry axis is proportional to the separation of the Dirac points and is independent of the microscopic model parameters.
This means that $\cso_{zz}$ reflects topological property of the electronic structure.
The spin-orbit crossed susceptibility is induced only along the rotational symmetry axis. We also calculate the spin susceptibility. The spin susceptibility is anisotropic and vanishingly small along the rotational symmetry axis, in contrast to the spin-orbit crossed susceptibility. The two susceptibilities are quantitatively compared for material parameters of ${\rm Cd}_2{\rm As}_3$ and ${\rm Na}_3{\rm Bi}$.
At the Dirac point, the orbital susceptibility logarithmically diverges and gives dominant contribution to the total susceptibility. Off the Dirac point, on the other hand, the orbital susceptibility decreases \cite{fukuyama1970interband,koshino2010anomalous,koshino2016magnetic}  and the contribution from the spin susceptibility and the spin-orbit crossed susceptibility is important for appropriate estimation of the total susceptibility.

\section*{ACNOWLEDGEMENT}
This work was supported by JSPS KAKENHI Grant Numbers JP15H05854 and JP17K05485, and JST CREST Grant Number JPMJCR18T2.

%\appendix
%
%\section{Analytical expressions}
%
%\begin{align}
%&|\la R,s,\bk|\s_z|R,s,\bk\ra|^2=\frac{(t^2\sin^2k_za+m_\bk^2-s\e t\sin k_za)^2}{\e^2(\e-st\sin k_za)^2}, \\
%&|\la L,s,\bk|\s_z|L,s,\bk\ra|^2=|\la R,s,\bk|\s_z|R,s,\bk\ra|^2, \\
%&|\la R,s,\bk|\s_z|L,s,\bk\ra|^2=\frac{t^2(\sin^2k_xa+\sin^2k_ya)m_\bk^2}{\e^2(\e-st\sin k_za)^2},
%\end{align}

\bibliography{TDS_SO_susceptibility}

\begin{thebibliography}{44}
\expandafter\ifx\csname natexlab\endcsname\relax\def\natexlab#1{#1}\fi
\expandafter\ifx\csname bibnamefont\endcsname\relax
  \def\bibnamefont#1{#1}\fi
\expandafter\ifx\csname bibfnamefont\endcsname\relax
  \def\bibfnamefont#1{#1}\fi
\expandafter\ifx\csname citenamefont\endcsname\relax
  \def\citenamefont#1{#1}\fi
\expandafter\ifx\csname url\endcsname\relax
  \def\url#1{\texttt{#1}}\fi
\expandafter\ifx\csname urlprefix\endcsname\relax\def\urlprefix{URL }\fi
\providecommand{\bibinfo}[2]{#2}
\providecommand{\eprint}[2][]{\url{#2}}

\bibitem[{\citenamefont{Hasan and Kane}(2010)}]{hasan2010topological}
\bibinfo{author}{\bibfnamefont{M.~Z.} \bibnamefont{Hasan}} \bibnamefont{and}
  \bibinfo{author}{\bibfnamefont{C.~L.} \bibnamefont{Kane}},
  \bibinfo{journal}{Rev. Mod. Phys.} \textbf{\bibinfo{volume}{82}},
  \bibinfo{pages}{3045} (\bibinfo{year}{2010}).

\bibitem[{\citenamefont{Armitage et~al.}(2018)\citenamefont{Armitage, Mele, and
  Vishwanath}}]{armitage2018weyl}
\bibinfo{author}{\bibfnamefont{N.~P.} \bibnamefont{Armitage}},
  \bibinfo{author}{\bibfnamefont{E.~J.} \bibnamefont{Mele}}, \bibnamefont{and}
  \bibinfo{author}{\bibfnamefont{A.}~\bibnamefont{Vishwanath}},
  \bibinfo{journal}{Rev. Mod. Phys.} \textbf{\bibinfo{volume}{90}},
  \bibinfo{pages}{015001} (\bibinfo{year}{2018}).

\bibitem[{\citenamefont{Yang and Chang}(2006)}]{yang2006streda}
\bibinfo{author}{\bibfnamefont{M.-F.} \bibnamefont{Yang}} \bibnamefont{and}
  \bibinfo{author}{\bibfnamefont{M.-C.} \bibnamefont{Chang}},
  \bibinfo{journal}{Phys. Rev. B} \textbf{\bibinfo{volume}{73}},
  \bibinfo{pages}{073304} (\bibinfo{year}{2006}).

\bibitem[{\citenamefont{Murakami}(2006)}]{murakami2006quantum}
\bibinfo{author}{\bibfnamefont{S.}~\bibnamefont{Murakami}},
  \bibinfo{journal}{Phys. Rev. Lett.} \textbf{\bibinfo{volume}{97}},
  \bibinfo{pages}{236805} (\bibinfo{year}{2006}).

\bibitem[{\citenamefont{Tserkovnyak et~al.}(2015)\citenamefont{Tserkovnyak,
  Pesin, and Loss}}]{tserkovnyak2015spin}
\bibinfo{author}{\bibfnamefont{Y.}~\bibnamefont{Tserkovnyak}},
  \bibinfo{author}{\bibfnamefont{D.~A.} \bibnamefont{Pesin}}, \bibnamefont{and}
  \bibinfo{author}{\bibfnamefont{D.}~\bibnamefont{Loss}},
  \bibinfo{journal}{Phys. Rev. B} \textbf{\bibinfo{volume}{91}},
  \bibinfo{pages}{041121} (\bibinfo{year}{2015}).

\bibitem[{\citenamefont{Koshino and Hizbullah}(2016)}]{koshino2016magnetic}
\bibinfo{author}{\bibfnamefont{M.}~\bibnamefont{Koshino}} \bibnamefont{and}
  \bibinfo{author}{\bibfnamefont{I.~F.} \bibnamefont{Hizbullah}},
  \bibinfo{journal}{Phys. Rev. B} \textbf{\bibinfo{volume}{93}},
  \bibinfo{pages}{045201} (\bibinfo{year}{2016}).

\bibitem[{\citenamefont{Nakai and Nomura}(2016)}]{nakai2016crossed}
\bibinfo{author}{\bibfnamefont{R.}~\bibnamefont{Nakai}} \bibnamefont{and}
  \bibinfo{author}{\bibfnamefont{K.}~\bibnamefont{Nomura}},
  \bibinfo{journal}{Phys. Rev. B} \textbf{\bibinfo{volume}{93}},
  \bibinfo{pages}{214434} (\bibinfo{year}{2016}).

\bibitem[{\citenamefont{Suzuura and Ando}(2016)}]{suzuura2016theory}
\bibinfo{author}{\bibfnamefont{H.}~\bibnamefont{Suzuura}} \bibnamefont{and}
  \bibinfo{author}{\bibfnamefont{T.}~\bibnamefont{Ando}},
  \bibinfo{journal}{Phys. Rev. B} \textbf{\bibinfo{volume}{94}},
  \bibinfo{pages}{085303} (\bibinfo{year}{2016}).

\bibitem[{\citenamefont{Ando and Suzuura}(2017)}]{ando2017theory}
\bibinfo{author}{\bibfnamefont{T.}~\bibnamefont{Ando}} \bibnamefont{and}
  \bibinfo{author}{\bibfnamefont{H.}~\bibnamefont{Suzuura}},
  \bibinfo{journal}{Journal of the Physical Society of Japan}
  \textbf{\bibinfo{volume}{86}}, \bibinfo{pages}{014701}
  (\bibinfo{year}{2017}).

\bibitem[{\citenamefont{Wang et~al.}(2012)\citenamefont{Wang, Sun, Chen,
  Franchini, Xu, Weng, Dai, and Fang}}]{wang2012dirac}
\bibinfo{author}{\bibfnamefont{Z.}~\bibnamefont{Wang}},
  \bibinfo{author}{\bibfnamefont{Y.}~\bibnamefont{Sun}},
  \bibinfo{author}{\bibfnamefont{X.-Q.} \bibnamefont{Chen}},
  \bibinfo{author}{\bibfnamefont{C.}~\bibnamefont{Franchini}},
  \bibinfo{author}{\bibfnamefont{G.}~\bibnamefont{Xu}},
  \bibinfo{author}{\bibfnamefont{H.}~\bibnamefont{Weng}},
  \bibinfo{author}{\bibfnamefont{X.}~\bibnamefont{Dai}}, \bibnamefont{and}
  \bibinfo{author}{\bibfnamefont{Z.}~\bibnamefont{Fang}},
  \bibinfo{journal}{Physical Review B} \textbf{\bibinfo{volume}{85}},
  \bibinfo{pages}{195320} (\bibinfo{year}{2012}).

\bibitem[{\citenamefont{Wang et~al.}(2013)\citenamefont{Wang, Weng, Wu, Dai,
  and Fang}}]{wang2013three}
\bibinfo{author}{\bibfnamefont{Z.}~\bibnamefont{Wang}},
  \bibinfo{author}{\bibfnamefont{H.}~\bibnamefont{Weng}},
  \bibinfo{author}{\bibfnamefont{Q.}~\bibnamefont{Wu}},
  \bibinfo{author}{\bibfnamefont{X.}~\bibnamefont{Dai}}, \bibnamefont{and}
  \bibinfo{author}{\bibfnamefont{Z.}~\bibnamefont{Fang}},
  \bibinfo{journal}{Physical Review B} \textbf{\bibinfo{volume}{88}},
  \bibinfo{pages}{125427} (\bibinfo{year}{2013}).

\bibitem[{\citenamefont{Morimoto and Furusaki}(2014)}]{morimoto2014weyl}
\bibinfo{author}{\bibfnamefont{T.}~\bibnamefont{Morimoto}} \bibnamefont{and}
  \bibinfo{author}{\bibfnamefont{A.}~\bibnamefont{Furusaki}},
  \bibinfo{journal}{Phys. Rev. B} \textbf{\bibinfo{volume}{89}},
  \bibinfo{pages}{235127} (\bibinfo{year}{2014}).

\bibitem[{\citenamefont{Yang and Nagaosa}(2014)}]{yang2014classification}
\bibinfo{author}{\bibfnamefont{B.-J.} \bibnamefont{Yang}} \bibnamefont{and}
  \bibinfo{author}{\bibfnamefont{N.}~\bibnamefont{Nagaosa}},
  \bibinfo{journal}{Nature Communications} \textbf{\bibinfo{volume}{5}},
  \bibinfo{pages}{4898 EP } (\bibinfo{year}{2014}), \bibinfo{note}{article}.

\bibitem[{\citenamefont{Pikulin et~al.}(2016)\citenamefont{Pikulin, Chen, and
  Franz}}]{pikulin2016chiral}
\bibinfo{author}{\bibfnamefont{D.}~\bibnamefont{Pikulin}},
  \bibinfo{author}{\bibfnamefont{A.}~\bibnamefont{Chen}}, \bibnamefont{and}
  \bibinfo{author}{\bibfnamefont{M.}~\bibnamefont{Franz}},
  \bibinfo{journal}{Physical Review X} \textbf{\bibinfo{volume}{6}},
  \bibinfo{pages}{041021} (\bibinfo{year}{2016}).

\bibitem[{\citenamefont{Cano et~al.}(2017)\citenamefont{Cano, Bradlyn, Wang,
  Hirschberger, Ong, and Bernevig}}]{cano2017chiral}
\bibinfo{author}{\bibfnamefont{J.}~\bibnamefont{Cano}},
  \bibinfo{author}{\bibfnamefont{B.}~\bibnamefont{Bradlyn}},
  \bibinfo{author}{\bibfnamefont{Z.}~\bibnamefont{Wang}},
  \bibinfo{author}{\bibfnamefont{M.}~\bibnamefont{Hirschberger}},
  \bibinfo{author}{\bibfnamefont{N.~P.} \bibnamefont{Ong}}, \bibnamefont{and}
  \bibinfo{author}{\bibfnamefont{B.~A.} \bibnamefont{Bernevig}},
  \bibinfo{journal}{Phys. Rev. B} \textbf{\bibinfo{volume}{95}},
  \bibinfo{pages}{161306} (\bibinfo{year}{2017}).

\bibitem[{\citenamefont{Liu et~al.}(2014)\citenamefont{Liu, Zhou, Zhang, Wang,
  Weng, Prabhakaran, Mo, Shen, Fang, Dai et~al.}}]{liu2014discovery}
\bibinfo{author}{\bibfnamefont{Z.}~\bibnamefont{Liu}},
  \bibinfo{author}{\bibfnamefont{B.}~\bibnamefont{Zhou}},
  \bibinfo{author}{\bibfnamefont{Y.}~\bibnamefont{Zhang}},
  \bibinfo{author}{\bibfnamefont{Z.}~\bibnamefont{Wang}},
  \bibinfo{author}{\bibfnamefont{H.}~\bibnamefont{Weng}},
  \bibinfo{author}{\bibfnamefont{D.}~\bibnamefont{Prabhakaran}},
  \bibinfo{author}{\bibfnamefont{S.-K.} \bibnamefont{Mo}},
  \bibinfo{author}{\bibfnamefont{Z.}~\bibnamefont{Shen}},
  \bibinfo{author}{\bibfnamefont{Z.}~\bibnamefont{Fang}},
  \bibinfo{author}{\bibfnamefont{X.}~\bibnamefont{Dai}}, \bibnamefont{et~al.},
  \bibinfo{journal}{Science} \textbf{\bibinfo{volume}{343}},
  \bibinfo{pages}{864} (\bibinfo{year}{2014}).

\bibitem[{\citenamefont{Neupane et~al.}(2014)\citenamefont{Neupane, Xu, Sankar,
  Alidoust, Bian, Liu, Belopolski, Chang, Jeng, Lin
  et~al.}}]{neupane2014observation}
\bibinfo{author}{\bibfnamefont{M.}~\bibnamefont{Neupane}},
  \bibinfo{author}{\bibfnamefont{S.-Y.} \bibnamefont{Xu}},
  \bibinfo{author}{\bibfnamefont{R.}~\bibnamefont{Sankar}},
  \bibinfo{author}{\bibfnamefont{N.}~\bibnamefont{Alidoust}},
  \bibinfo{author}{\bibfnamefont{G.}~\bibnamefont{Bian}},
  \bibinfo{author}{\bibfnamefont{C.}~\bibnamefont{Liu}},
  \bibinfo{author}{\bibfnamefont{I.}~\bibnamefont{Belopolski}},
  \bibinfo{author}{\bibfnamefont{T.-R.} \bibnamefont{Chang}},
  \bibinfo{author}{\bibfnamefont{H.-T.} \bibnamefont{Jeng}},
  \bibinfo{author}{\bibfnamefont{H.}~\bibnamefont{Lin}}, \bibnamefont{et~al.},
  \bibinfo{journal}{Nature communications} \textbf{\bibinfo{volume}{5}}
  (\bibinfo{year}{2014}).

\bibitem[{\citenamefont{Borisenko et~al.}(2014)\citenamefont{Borisenko, Gibson,
  Evtushinsky, Zabolotnyy, B{\"u}chner, and Cava}}]{borisenko2014experimental}
\bibinfo{author}{\bibfnamefont{S.}~\bibnamefont{Borisenko}},
  \bibinfo{author}{\bibfnamefont{Q.}~\bibnamefont{Gibson}},
  \bibinfo{author}{\bibfnamefont{D.}~\bibnamefont{Evtushinsky}},
  \bibinfo{author}{\bibfnamefont{V.}~\bibnamefont{Zabolotnyy}},
  \bibinfo{author}{\bibfnamefont{B.}~\bibnamefont{B{\"u}chner}},
  \bibnamefont{and} \bibinfo{author}{\bibfnamefont{R.~J.} \bibnamefont{Cava}},
  \bibinfo{journal}{Physical review letters} \textbf{\bibinfo{volume}{113}},
  \bibinfo{pages}{027603} (\bibinfo{year}{2014}).

\bibitem[{\citenamefont{Murakami}(2007)}]{murakami2007phase}
\bibinfo{author}{\bibfnamefont{S.}~\bibnamefont{Murakami}},
  \bibinfo{journal}{New Journal of Physics} \textbf{\bibinfo{volume}{9}},
  \bibinfo{pages}{356} (\bibinfo{year}{2007}).

\bibitem[{\citenamefont{Guo et~al.}(2011)\citenamefont{Guo, Sugawara, Takayama,
  Souma, Sato, Satoh, Ohnishi, Kitaura, Sasaki, Xue et~al.}}]{guo2011evolution}
\bibinfo{author}{\bibfnamefont{H.}~\bibnamefont{Guo}},
  \bibinfo{author}{\bibfnamefont{K.}~\bibnamefont{Sugawara}},
  \bibinfo{author}{\bibfnamefont{A.}~\bibnamefont{Takayama}},
  \bibinfo{author}{\bibfnamefont{S.}~\bibnamefont{Souma}},
  \bibinfo{author}{\bibfnamefont{T.}~\bibnamefont{Sato}},
  \bibinfo{author}{\bibfnamefont{N.}~\bibnamefont{Satoh}},
  \bibinfo{author}{\bibfnamefont{A.}~\bibnamefont{Ohnishi}},
  \bibinfo{author}{\bibfnamefont{M.}~\bibnamefont{Kitaura}},
  \bibinfo{author}{\bibfnamefont{M.}~\bibnamefont{Sasaki}},
  \bibinfo{author}{\bibfnamefont{Q.-K.} \bibnamefont{Xue}},
  \bibnamefont{et~al.}, \bibinfo{journal}{Physical Review B}
  \textbf{\bibinfo{volume}{83}}, \bibinfo{pages}{201104}
  (\bibinfo{year}{2011}).

\bibitem[{\citenamefont{Xu et~al.}(2011)\citenamefont{Xu, Xia, Wray, Jia,
  Meier, Dil, Osterwalder, Slomski, Bansil, Lin et~al.}}]{xu2011topological}
\bibinfo{author}{\bibfnamefont{S.-Y.} \bibnamefont{Xu}},
  \bibinfo{author}{\bibfnamefont{Y.}~\bibnamefont{Xia}},
  \bibinfo{author}{\bibfnamefont{L.}~\bibnamefont{Wray}},
  \bibinfo{author}{\bibfnamefont{S.}~\bibnamefont{Jia}},
  \bibinfo{author}{\bibfnamefont{F.}~\bibnamefont{Meier}},
  \bibinfo{author}{\bibfnamefont{J.}~\bibnamefont{Dil}},
  \bibinfo{author}{\bibfnamefont{J.}~\bibnamefont{Osterwalder}},
  \bibinfo{author}{\bibfnamefont{B.}~\bibnamefont{Slomski}},
  \bibinfo{author}{\bibfnamefont{A.}~\bibnamefont{Bansil}},
  \bibinfo{author}{\bibfnamefont{H.}~\bibnamefont{Lin}}, \bibnamefont{et~al.},
  \bibinfo{journal}{Science} \textbf{\bibinfo{volume}{332}},
  \bibinfo{pages}{560} (\bibinfo{year}{2011}).

\bibitem[{\citenamefont{Sato et~al.}(2011)\citenamefont{Sato, Segawa, Kosaka,
  Souma, Nakayama, Eto, Minami, Ando, and Takahashi}}]{sato2011unexpected}
\bibinfo{author}{\bibfnamefont{T.}~\bibnamefont{Sato}},
  \bibinfo{author}{\bibfnamefont{K.}~\bibnamefont{Segawa}},
  \bibinfo{author}{\bibfnamefont{K.}~\bibnamefont{Kosaka}},
  \bibinfo{author}{\bibfnamefont{S.}~\bibnamefont{Souma}},
  \bibinfo{author}{\bibfnamefont{K.}~\bibnamefont{Nakayama}},
  \bibinfo{author}{\bibfnamefont{K.}~\bibnamefont{Eto}},
  \bibinfo{author}{\bibfnamefont{T.}~\bibnamefont{Minami}},
  \bibinfo{author}{\bibfnamefont{Y.}~\bibnamefont{Ando}}, \bibnamefont{and}
  \bibinfo{author}{\bibfnamefont{T.}~\bibnamefont{Takahashi}},
  \bibinfo{journal}{Nature Physics} \textbf{\bibinfo{volume}{7}},
  \bibinfo{pages}{840} (\bibinfo{year}{2011}).

\bibitem[{\citenamefont{Burkov and Kim}(2016)}]{burkov2016z2}
\bibinfo{author}{\bibfnamefont{A.~A.} \bibnamefont{Burkov}} \bibnamefont{and}
  \bibinfo{author}{\bibfnamefont{Y.~B.} \bibnamefont{Kim}},
  \bibinfo{journal}{Phys. Rev. Lett.} \textbf{\bibinfo{volume}{117}},
  \bibinfo{pages}{136602} (\bibinfo{year}{2016}).

\bibitem[{\citenamefont{Fukuyama and Kubo}(1970)}]{fukuyama1970interband}
\bibinfo{author}{\bibfnamefont{H.}~\bibnamefont{Fukuyama}} \bibnamefont{and}
  \bibinfo{author}{\bibfnamefont{R.}~\bibnamefont{Kubo}},
  \bibinfo{journal}{Journal of the Physical Society of Japan}
  \textbf{\bibinfo{volume}{28}}, \bibinfo{pages}{570} (\bibinfo{year}{1970}).

\bibitem[{\citenamefont{Koshino and Ando}(2010)}]{koshino2010anomalous}
\bibinfo{author}{\bibfnamefont{M.}~\bibnamefont{Koshino}} \bibnamefont{and}
  \bibinfo{author}{\bibfnamefont{T.}~\bibnamefont{Ando}},
  \bibinfo{journal}{Phys. Rev. B} \textbf{\bibinfo{volume}{81}},
  \bibinfo{pages}{195431} (\bibinfo{year}{2010}).

\bibitem[{\citenamefont{Mikitik and Sharlai}(2016)}]{mikitik2016magnetic}
\bibinfo{author}{\bibfnamefont{G.~P.} \bibnamefont{Mikitik}} \bibnamefont{and}
  \bibinfo{author}{\bibfnamefont{Y.~V.} \bibnamefont{Sharlai}},
  \bibinfo{journal}{Phys. Rev. B} \textbf{\bibinfo{volume}{94}},
  \bibinfo{pages}{195123} (\bibinfo{year}{2016}).

\bibitem[{\citenamefont{Thakur et~al.}(2018)\citenamefont{Thakur, Sadhukhan,
  and Agarwal}}]{thakur2018dynamic}
\bibinfo{author}{\bibfnamefont{A.}~\bibnamefont{Thakur}},
  \bibinfo{author}{\bibfnamefont{K.}~\bibnamefont{Sadhukhan}},
  \bibnamefont{and} \bibinfo{author}{\bibfnamefont{A.}~\bibnamefont{Agarwal}},
  \bibinfo{journal}{Phys. Rev. B} \textbf{\bibinfo{volume}{97}},
  \bibinfo{pages}{035403} (\bibinfo{year}{2018}).

\bibitem[{\citenamefont{Zhou and Chang}(2018)}]{zhou2018dynamical}
\bibinfo{author}{\bibfnamefont{J.}~\bibnamefont{Zhou}} \bibnamefont{and}
  \bibinfo{author}{\bibfnamefont{H.-R.} \bibnamefont{Chang}},
  \bibinfo{journal}{Phys. Rev. B} \textbf{\bibinfo{volume}{97}},
  \bibinfo{pages}{075202} (\bibinfo{year}{2018}).

\bibitem[{\citenamefont{Ominato and Nomura}(2018)}]{ominato2018spin}
\bibinfo{author}{\bibfnamefont{Y.}~\bibnamefont{Ominato}} \bibnamefont{and}
  \bibinfo{author}{\bibfnamefont{K.}~\bibnamefont{Nomura}},
  \bibinfo{journal}{Phys. Rev. B} \textbf{\bibinfo{volume}{97}},
  \bibinfo{pages}{245207} (\bibinfo{year}{2018}).

\bibitem[{\citenamefont{Van~Vleck}(1932)}]{van1932theory}
\bibinfo{author}{\bibfnamefont{J.}~\bibnamefont{Van~Vleck}},
  \emph{\bibinfo{title}{The theory of electronic and magnetic susceptibility}}
  (\bibinfo{year}{1932}).

\bibitem[{\citenamefont{Yu et~al.}(2010)\citenamefont{Yu, Zhang, Zhang, Zhang,
  Dai, and Fang}}]{yu2010quantized}
\bibinfo{author}{\bibfnamefont{R.}~\bibnamefont{Yu}},
  \bibinfo{author}{\bibfnamefont{W.}~\bibnamefont{Zhang}},
  \bibinfo{author}{\bibfnamefont{H.-J.} \bibnamefont{Zhang}},
  \bibinfo{author}{\bibfnamefont{S.-C.} \bibnamefont{Zhang}},
  \bibinfo{author}{\bibfnamefont{X.}~\bibnamefont{Dai}}, \bibnamefont{and}
  \bibinfo{author}{\bibfnamefont{Z.}~\bibnamefont{Fang}},
  \bibinfo{journal}{Science} \textbf{\bibinfo{volume}{329}},
  \bibinfo{pages}{61} (\bibinfo{year}{2010}).

\bibitem[{\citenamefont{Zhang et~al.}(2013)\citenamefont{Zhang, Chang, Tang,
  Zhang, Feng, Li, Wang, Chen, Liu, Duan et~al.}}]{zhang2013topology}
\bibinfo{author}{\bibfnamefont{J.}~\bibnamefont{Zhang}},
  \bibinfo{author}{\bibfnamefont{C.-Z.} \bibnamefont{Chang}},
  \bibinfo{author}{\bibfnamefont{P.}~\bibnamefont{Tang}},
  \bibinfo{author}{\bibfnamefont{Z.}~\bibnamefont{Zhang}},
  \bibinfo{author}{\bibfnamefont{X.}~\bibnamefont{Feng}},
  \bibinfo{author}{\bibfnamefont{K.}~\bibnamefont{Li}},
  \bibinfo{author}{\bibfnamefont{L.-l.} \bibnamefont{Wang}},
  \bibinfo{author}{\bibfnamefont{X.}~\bibnamefont{Chen}},
  \bibinfo{author}{\bibfnamefont{C.}~\bibnamefont{Liu}},
  \bibinfo{author}{\bibfnamefont{W.}~\bibnamefont{Duan}}, \bibnamefont{et~al.},
  \bibinfo{journal}{Science} \textbf{\bibinfo{volume}{339}},
  \bibinfo{pages}{1582} (\bibinfo{year}{2013}).

\bibitem[{\citenamefont{Bernevig et~al.}(2006)\citenamefont{Bernevig, Hughes,
  and Zhang}}]{bernevig2006quantum}
\bibinfo{author}{\bibfnamefont{B.~A.} \bibnamefont{Bernevig}},
  \bibinfo{author}{\bibfnamefont{T.~L.} \bibnamefont{Hughes}},
  \bibnamefont{and} \bibinfo{author}{\bibfnamefont{S.-C.} \bibnamefont{Zhang}},
  \bibinfo{journal}{Science} \textbf{\bibinfo{volume}{314}},
  \bibinfo{pages}{1757} (\bibinfo{year}{2006}).

\bibitem[{\citenamefont{Liu et~al.}(2010)\citenamefont{Liu, Qi, Zhang, Dai,
  Fang, and Zhang}}]{liu2010model}
\bibinfo{author}{\bibfnamefont{C.-X.} \bibnamefont{Liu}},
  \bibinfo{author}{\bibfnamefont{X.-L.} \bibnamefont{Qi}},
  \bibinfo{author}{\bibfnamefont{H.}~\bibnamefont{Zhang}},
  \bibinfo{author}{\bibfnamefont{X.}~\bibnamefont{Dai}},
  \bibinfo{author}{\bibfnamefont{Z.}~\bibnamefont{Fang}}, \bibnamefont{and}
  \bibinfo{author}{\bibfnamefont{S.-C.} \bibnamefont{Zhang}},
  \bibinfo{journal}{Physical Review B} \textbf{\bibinfo{volume}{82}},
  \bibinfo{pages}{045122} (\bibinfo{year}{2010}).

\bibitem[{\citenamefont{Wakatsuki et~al.}(2015)\citenamefont{Wakatsuki, Ezawa,
  and Nagaosa}}]{wakatsuki2015domain}
\bibinfo{author}{\bibfnamefont{R.}~\bibnamefont{Wakatsuki}},
  \bibinfo{author}{\bibfnamefont{M.}~\bibnamefont{Ezawa}}, \bibnamefont{and}
  \bibinfo{author}{\bibfnamefont{N.}~\bibnamefont{Nagaosa}},
  \bibinfo{journal}{Scientific Reports} \textbf{\bibinfo{volume}{5}},
  \bibinfo{pages}{13638 EP } (\bibinfo{year}{2015}).

\bibitem[{\citenamefont{Sundaram and Niu}(1999)}]{sundarm1999wave}
\bibinfo{author}{\bibfnamefont{G.}~\bibnamefont{Sundaram}} \bibnamefont{and}
  \bibinfo{author}{\bibfnamefont{Q.}~\bibnamefont{Niu}},
  \bibinfo{journal}{Phys. Rev. B} \textbf{\bibinfo{volume}{59}},
  \bibinfo{pages}{14915} (\bibinfo{year}{1999}).

\bibitem[{\citenamefont{Xiao et~al.}(2005)\citenamefont{Xiao, Shi, and
  Niu}}]{xiao2005berry}
\bibinfo{author}{\bibfnamefont{D.}~\bibnamefont{Xiao}},
  \bibinfo{author}{\bibfnamefont{J.}~\bibnamefont{Shi}}, \bibnamefont{and}
  \bibinfo{author}{\bibfnamefont{Q.}~\bibnamefont{Niu}},
  \bibinfo{journal}{Phys. Rev. Lett.} \textbf{\bibinfo{volume}{95}},
  \bibinfo{pages}{137204} (\bibinfo{year}{2005}).

\bibitem[{\citenamefont{Thonhauser et~al.}(2005)\citenamefont{Thonhauser,
  Ceresoli, Vanderbilt, and Resta}}]{thonhauser2005orbital}
\bibinfo{author}{\bibfnamefont{T.}~\bibnamefont{Thonhauser}},
  \bibinfo{author}{\bibfnamefont{D.}~\bibnamefont{Ceresoli}},
  \bibinfo{author}{\bibfnamefont{D.}~\bibnamefont{Vanderbilt}},
  \bibnamefont{and} \bibinfo{author}{\bibfnamefont{R.}~\bibnamefont{Resta}},
  \bibinfo{journal}{Phys. Rev. Lett.} \textbf{\bibinfo{volume}{95}},
  \bibinfo{pages}{137205} (\bibinfo{year}{2005}).

\bibitem[{\citenamefont{Ceresoli et~al.}(2006)\citenamefont{Ceresoli,
  Thonhauser, Vanderbilt, and Resta}}]{ceresoli2006orbital}
\bibinfo{author}{\bibfnamefont{D.}~\bibnamefont{Ceresoli}},
  \bibinfo{author}{\bibfnamefont{T.}~\bibnamefont{Thonhauser}},
  \bibinfo{author}{\bibfnamefont{D.}~\bibnamefont{Vanderbilt}},
  \bibnamefont{and} \bibinfo{author}{\bibfnamefont{R.}~\bibnamefont{Resta}},
  \bibinfo{journal}{Phys. Rev. B} \textbf{\bibinfo{volume}{74}},
  \bibinfo{pages}{024408} (\bibinfo{year}{2006}).

\bibitem[{\citenamefont{Shi et~al.}(2007)\citenamefont{Shi, Vignale, Xiao, and
  Niu}}]{shi2007quantum}
\bibinfo{author}{\bibfnamefont{J.}~\bibnamefont{Shi}},
  \bibinfo{author}{\bibfnamefont{G.}~\bibnamefont{Vignale}},
  \bibinfo{author}{\bibfnamefont{D.}~\bibnamefont{Xiao}}, \bibnamefont{and}
  \bibinfo{author}{\bibfnamefont{Q.}~\bibnamefont{Niu}},
  \bibinfo{journal}{Phys. Rev. Lett.} \textbf{\bibinfo{volume}{99}},
  \bibinfo{pages}{197202} (\bibinfo{year}{2007}).

\bibitem[{\citenamefont{Streda}(1982)}]{streda1982theory}
\bibinfo{author}{\bibfnamefont{P.}~\bibnamefont{Streda}},
  \bibinfo{journal}{Journal of Physics C: Solid State Physics}
  \textbf{\bibinfo{volume}{15}}, \bibinfo{pages}{L717} (\bibinfo{year}{1982}).

\bibitem[{\citenamefont{Burkov and Balents}(2011)}]{burkov2011weyl}
\bibinfo{author}{\bibfnamefont{A.~A.} \bibnamefont{Burkov}} \bibnamefont{and}
  \bibinfo{author}{\bibfnamefont{L.}~\bibnamefont{Balents}},
  \bibinfo{journal}{Phys. Rev. Lett.} \textbf{\bibinfo{volume}{107}},
  \bibinfo{pages}{127205} (\bibinfo{year}{2011}).

\bibitem[{\citenamefont{Jeon et~al.}(2014)\citenamefont{Jeon, Zhou, Gyenis,
  Feldman, Kimchi, Potter, Gibson, Cava, Vishwanath, and
  Yazdani}}]{jeon2014landau}
\bibinfo{author}{\bibfnamefont{S.}~\bibnamefont{Jeon}},
  \bibinfo{author}{\bibfnamefont{B.~B.} \bibnamefont{Zhou}},
  \bibinfo{author}{\bibfnamefont{A.}~\bibnamefont{Gyenis}},
  \bibinfo{author}{\bibfnamefont{B.~E.} \bibnamefont{Feldman}},
  \bibinfo{author}{\bibfnamefont{I.}~\bibnamefont{Kimchi}},
  \bibinfo{author}{\bibfnamefont{A.~C.} \bibnamefont{Potter}},
  \bibinfo{author}{\bibfnamefont{Q.~D.} \bibnamefont{Gibson}},
  \bibinfo{author}{\bibfnamefont{R.~J.} \bibnamefont{Cava}},
  \bibinfo{author}{\bibfnamefont{A.}~\bibnamefont{Vishwanath}},
  \bibnamefont{and} \bibinfo{author}{\bibfnamefont{A.}~\bibnamefont{Yazdani}},
  \bibinfo{journal}{Nature Materials} \textbf{\bibinfo{volume}{13}},
  \bibinfo{pages}{851 EP } (\bibinfo{year}{2014}).

\bibitem[{\citenamefont{Xiong et~al.}(2015)\citenamefont{Xiong, Kushwaha,
  Liang, Krizan, Hirschberger, Wang, Cava, and Ong}}]{xiong2015evidence}
\bibinfo{author}{\bibfnamefont{J.}~\bibnamefont{Xiong}},
  \bibinfo{author}{\bibfnamefont{S.~K.} \bibnamefont{Kushwaha}},
  \bibinfo{author}{\bibfnamefont{T.}~\bibnamefont{Liang}},
  \bibinfo{author}{\bibfnamefont{J.~W.} \bibnamefont{Krizan}},
  \bibinfo{author}{\bibfnamefont{M.}~\bibnamefont{Hirschberger}},
  \bibinfo{author}{\bibfnamefont{W.}~\bibnamefont{Wang}},
  \bibinfo{author}{\bibfnamefont{R.}~\bibnamefont{Cava}}, \bibnamefont{and}
  \bibinfo{author}{\bibfnamefont{N.}~\bibnamefont{Ong}},
  \bibinfo{journal}{Science} \textbf{\bibinfo{volume}{350}},
  \bibinfo{pages}{413} (\bibinfo{year}{2015}).

\end{thebibliography}

\end{document}